\documentclass[twocolumn,showpacs,amsmath,amssymb,superscriptaddress,prb]{revtex4}

\usepackage{graphicx}
\usepackage{dcolumn}
\usepackage{bm}

\newcommand{\DTTO}{Dy$_{x}$Tb$_{2-x}$Ti$_{2}$O$_{7}$}
\newcommand{\chireal}{$\chi^{\prime}(T)$}
\newcommand{\chiimag}{$\chi^{\prime\prime}(T)$}
\newcommand{\Ts}{$T_{s}$}
\newcommand{\Tstar}{$T^{*}$}
\newcommand{\Tf}{$T_{f}$}
\newcommand{\Dyion}{Dy$^{3+}$}
\newcommand{\Tbion}{Tb$^{3+}$}
\newcommand{\Ea}{$E_{A}$}

\begin{document}

\title{Emergent order in the spin-frustrated system Dy$_{x}$Tb$_{2-x}$Ti$_{2}$O$_{7}$ studied by ac susceptibility measurements}
\author{Hui Xing}
\affiliation{Department of Physics, Zhejiang University, Hangzhou 310027, China}
\affiliation{Department of Physics, University at Buffalo, the State University of New York, Buffalo, New York 14260, USA}
\author{Mi He}
\affiliation{Department of Physics, Zhejiang University, Hangzhou 310027, China}
\author{Chunmu Feng}
\affiliation{Test and Analysis Center, Zhejiang University,
Hangzhou 310027, China}
\author{Hanjie Guo}
\affiliation{Department of Physics, Zhejiang University, Hangzhou 310027, China}
\author{Hao Zeng}
\affiliation{Department of Physics, University at Buffalo, the State University of New York, Buffalo, New York 14260, USA}
\author{Zhu-An Xu}
\email{zhuan@zju.edu.cn} \affiliation{Department of Physics, Zhejiang University, Hangzhou 310027, China}

\begin{abstract}
We report the a.c. susceptibility study of \DTTO\ with $x\in$ {[}0,
2{]}. In addition to the single-ion effect at $T_{s}$ (single-ion
effect peak temperature) corresponding to the Dy$^{3+}$ spins as
that in spin ice Dy$_{2}$Ti$_{2}$O$_{7}$ and a possible spin
freezing peak at $T_{f}$ ($T_{f} < 3$ K), a new peak associated with
\Tbion\ is observed in $\chi_{ac}(T)$ at nonzero magnetic field with
a characteristic temperature \Tstar\ ($T_{f} < T^{*} < T_{s}$). \Tstar\
increases linearly with $x$ in a wide composition range ($0 < x <
1.5$ at 5 kOe). Both application of a magnetic field and increasing
doping with \Dyion\ enhance \Tstar. The \Tstar\ peak is found to be
thermally driven with an unusually large energy barrier as indicated
from its frequency dependence. These effects are closely related to
the crystal field levels, and the underlying mechanism remains to be
understood.

\end{abstract}

\maketitle

\section{introduction}
Geometrically frustrated magnetic materials,\cite{Ramirez1994,
Schiffer1996, Diep2005} which are ordered systems with specific
geometry that accommodates closely competing spin-spin interactions
and therefore frustration, attract extensive interests because of
their novel ground states. The pyrochlores \cite{Gardner2009}
provide a variety of geometrically frustrated magnetic materials,
including spin liquid,\cite{Gardner1999} spin ice,\cite{Harris1997,
Bramwell2001} and spin glass,\cite{Gingras1997-SG, Gardner1999-SG,
Greedan2009-SG} where magnetic exchange and dipolar interactions of
the nearest-neighbor spins dominate the low-temperature magnetic
property. Among these, two pyrochlores are of particular interests,
spin ice Dy$_{2}$Ti$_{2}$O$_{7}$ (DTO), Ho$_{2}$Ti$_{2}$O$_{7}$
and spin liquid Tb$_{2}$Ti$_{2}$O$_{7}$ (TTO), which have completely
different ground states while sharing essential similarities.

For the spin ice compound, spins upon each tetrahedron adopt the
two-in-two-out configuration in the ground state, which is a direct
analogy to the two-short-two-long proton bond configuration in water
ice.\cite{Harris1997} This results in the zero point
entropy.\cite{Ramirez1999} Strong crystal field (CF) splitting
induces uniaxial anisotropy of \Dyion\ spins giving rise to the
Ising-like ground state doublet, in DTO, with spins pointing along
the local (111) axis, which lies more than 200 K below the first
excited state.\cite{Rosenkranz2000, Matsuhira2001, Snyder2001} Novel
spin dynamics has been observed. The single-ion effect at 15 K and
spin freezing into the ice state\cite{Snyder2003, Snyder2004,
Ke2007, Snyder2004b} (strictly, spins falling into compliance with
the ice rule and freezing into disordered states) below 2 K are
found to be responsible for the low-temperature dynamic properties.
Moreover, it has been predicted\cite{Castelnovo2008, Jaubert2009}
and most recently verified experimentally\cite{Fennell2009,
Bramwell2009, Kadowaki2009} that the dipole-dipole interactions give
rise to magnetic monopoles, therefore a new wave of reconsideration
of the equivalency of electricity and magnetism is invoked.

The spin liquid compound TTO has Ising type spins similar to DTO but
with a much smaller gap of 18.7 K between the ground and excited
spin states, \cite{Gingras2000prb, Gardner2001} which is one order
of magnitude smaller than that of DTO. TTO remains in spin liquid
state down to 50 mK despite a short-range antiferromagnetic (AFM)
order.\cite{Gardner2001, Gardner2003-50mk} With a Curies-Weiss
temperature of -14 K (Ref.\cite{Gingras2000prb}), the absence of an
ordered ground state is unusual and has been attributed to the
delicate balance between the AFM exchange and ferromagnetic (FM)
dipolar interactions, which then places TTO right at the phase
boundary according to the phase diagram by Hertog and
Gingras.\cite{Hertog2000} This delicate balance, however, is
vulnerable to external perturbations, such as pressure and magnetic
fields. For example, TTO shows an AFM ground state with
spin-liquid-like fluctuations under high
pressure,\cite{Mirebeau2002} a long range order with spin wave
excitations \cite{Rule2006} and a spin-ice-like order with the $k =
0$ propagation vector \cite{Cao2008} under high magnetic fields
along the {[}110{]} axis. It is also proposed that quantum
fluctuations are responsible for the lack of an ordered ground
state, and TTO is argued to be in a quantum mechanically fluctuating
spin ice state.\cite{Molavian2007, Molavian2009}

It is therefore interesting to ask how the ground state and magnetic
interactions evolve if both Dy and Tb ions are introduced into the
rare earth (RE) sites. In a brief report\cite{Chang2007}, both spin
ice and liquid signatures are found in the ground state of
Ho$_{x}$Tb$_{2-x}$Ti$_{2}$O$_{7}$. Here, using the a.c.
susceptibility study on \DTTO\ (DTTO), we observed a new peak in
$\chi_{ac}(T)$ at the characteristic temperature $T^{*}$ ($T_{f} <
T^{*} < T_{s}$). \Tstar\ is found to increase linearly with Dy
composition $x$ in a surprisingly wide range ($0 < x < 1.5$) for $H
= 5$ kOe. The phase diagram of the composition dependent \Tstar\ is
obtained. The \Tstar\ peak is found to be thermally driven with an
unusually large energy barrier as indicated from its frequency
dependence. The observed peaks are closely related to the CF, and
possible origins are discussed using the composition and the field
dependence of a.c. susceptibility.\cite{notes-0}

\begin{figure}[t]
\includegraphics[width=0.95\columnwidth]{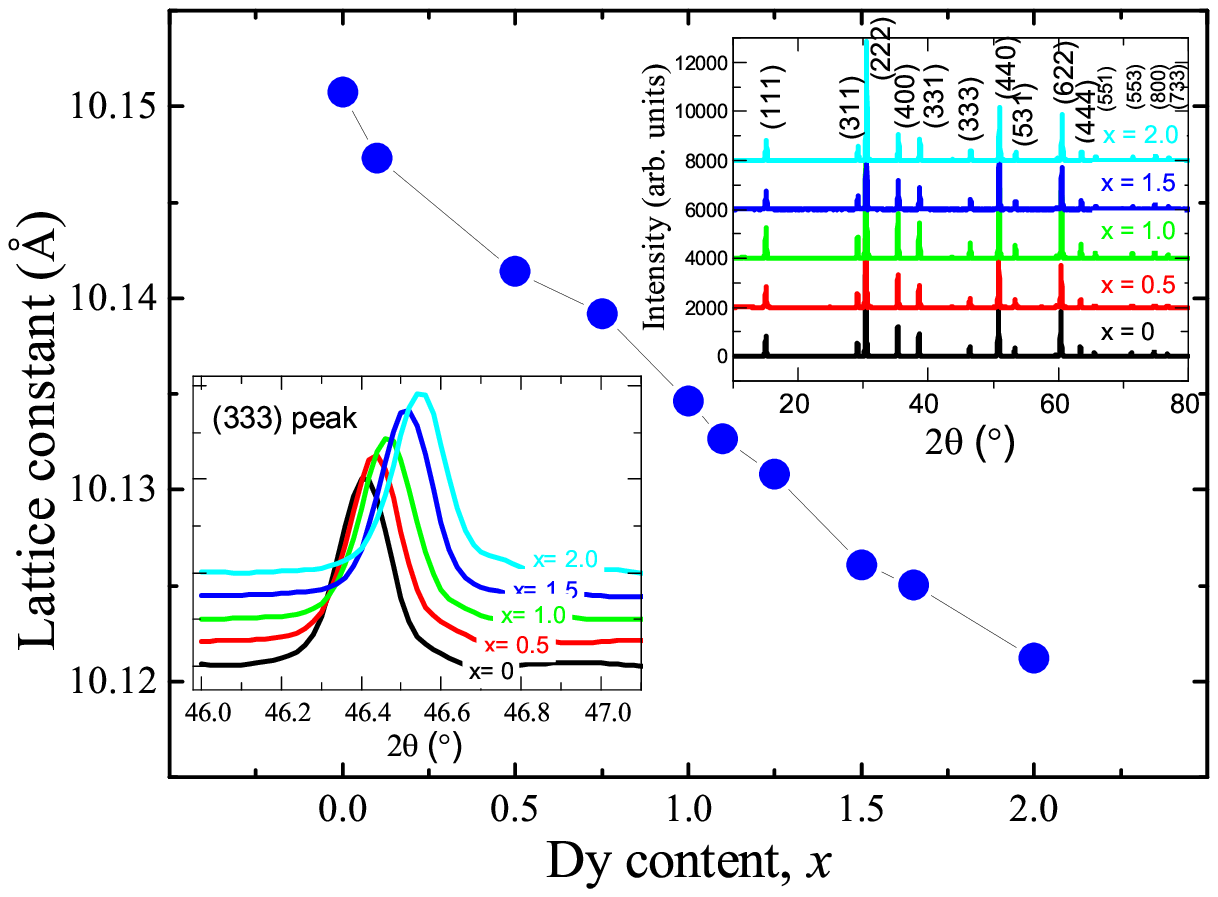}
\caption{\label{xrd}(color online)  The lattice constant (cubic
phase, a = b = c) of of \DTTO\ as a function of $x$ show a
quasi-linear decrease. (Upper inset) X-Ray diffractions show single
phase (pyrochlore structure) of all samples. (lower inset) Expanded
XRD patterns show the peak shift with $x$(Dy), the full width at
half maximum does not show increase, demonstrating the absence of
phase separation.}
\end{figure}

\section{experimental methods}

Polycrystalline DTTO with $x \in {[}0, 2{]}$ were prepared by the
standard solid-state reaction method.\cite{Snyder2001, Gardner1999}.
The X-ray diffraction data show that all samples are of single phase
with a pyrochlore structure. The lattice constants (cubic phase, $a
= b = c$) are shown in Fig. \ref{xrd}. For the two samples at the
composition boundaries, lattice constants of 10.15(1) \AA\ for TTO
($x$ = 2.0) and 10.12(1) \AA\ for DTO ($x$ = 0) are in good
agreement with previous reports,\cite{Gardner1999, Diep2005}
Curie-Weiss temperatures by fitting to the d.c. susceptibility between
10 and 20 K are consistent with values reported.\cite{notes-0} As there
are two spin species on the RE sites, it is important to address the
issue of possible phase separation. As shown in the lower inset of
Fig. \ref{xrd}, the full width at half maximum of the diffraction
peaks does not show obvious increase with different $x$, which
indicates that the system is free from phase separations. The
quasi-linear decrease of the lattice constants with increasing $x$
further confirms the samples' quality. D.c. and a.c.
susceptibilities are measured by a Quantum Design SQUID magnetometer
and a physical properties measurement system (PPMS) using an excitation field of $H_{ac} = 10$ Oe at frequency $f$ $(10 \leqslant\ f\leqslant\ 10 000$ Hz) respectively.

\section{results and discussions}
\begin{figure}[t]
 \includegraphics[width=0.95\columnwidth]{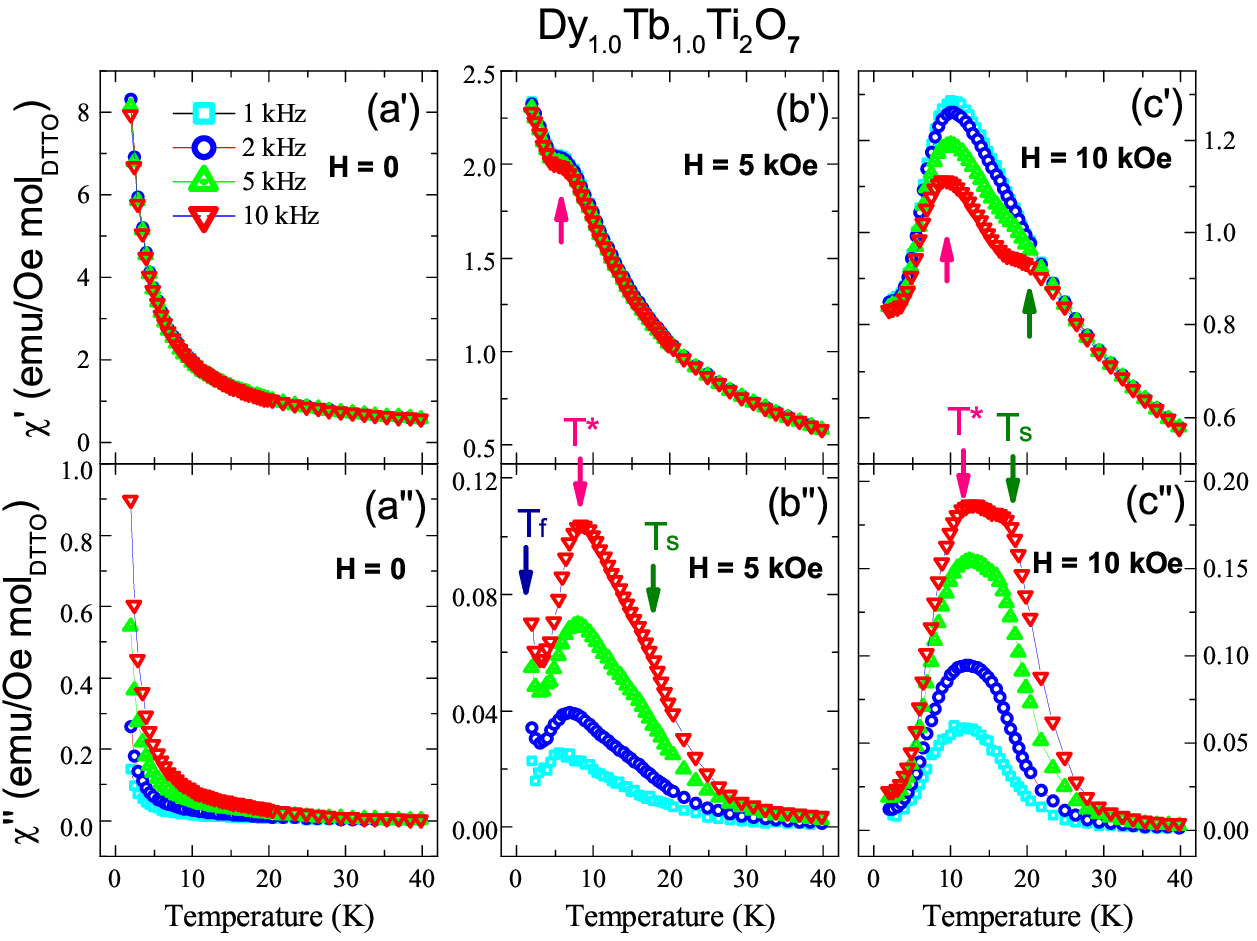}
\caption{\label{field-dependence}(color online) The real (upper
panel) and imaginary part (lower panel) of a.c. susceptibility of
Dy$_{1.0}$Tb$_{1.0}$Ti$_{2}$O$_{7}$ for several fields. Pointed by
the arrows are: \Tf\ peak (blue arrow), \Tstar\ peak (red arrow)
and \Ts\ peak (green arrow) denoting the low temperature freezing peak, the peak associated with \Tbion, and the single-ion peak associated with \Dyion\ respectively, as discussed in Fig.\ref{x-dependence} and corresponding text.}
\end{figure}

In Fig. \ref{field-dependence}, we take sample $x = 1.0$ as an
example and show the real and imaginary parts of the a.c.
susceptibility at varying magnetic fields. At $H = 0$, \chireal\
increases monotonically with an increasing slope upon cooling for
all excitation frequencies in study ($f$ = 1, 2, 5, 10 kHz), which
signals a canonical paramagnetic behavior.\cite{Ueland2006,Ke2007} The
imaginary part \chiimag\ in the lower panel shows a sharp increase
as well, even when approaching our lowest available temperature 2 K.
Therefore there should exist a strongly dissipative process at $T <
2$ K. The zero field a.c. susceptibility is qualitatively similar to
that of TTO (Ref.\cite{Ueland2006}) for all DTTO samples in study.
The magnitude of \chireal\ at $H = 0$ grows with increasing $x$,
presumably due to the larger moment of \Dyion\ spins. As the field
increases to 5 kOe, a clear dip in \chireal\ appears around 8.5 K,
leading to a local maximum in \chireal\ as well as a correlated peak
in \chiimag\ dictated by the Kramers-Kronig relation. It is also
notable that there is an emerging maximum on the right shoulder of
the peak in \chiimag, which becomes prominent as the field
increases. At $H = 10$ kOe, \chireal\ shows two clear dips at 12 K
and 17.5 K, thereby two local maxima appear in \chiimag. As the
frequency increases, the separation between the two local maxima
increases (Fig.\ref{field-dependence} c$^{\prime\prime}$).

\begin{figure*}[t]
\centering \includegraphics[width=1.9\columnwidth]{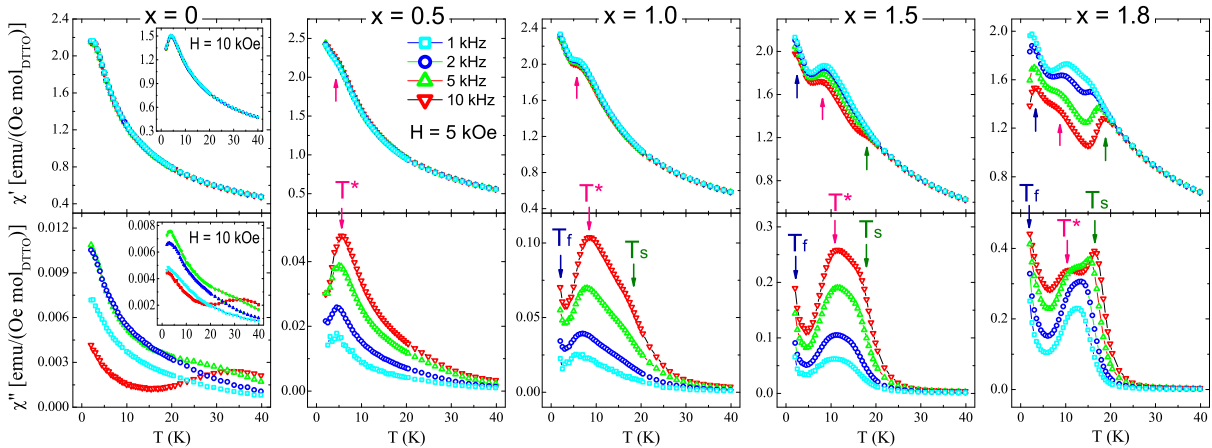}

\caption{\label{x-dependence}(color online) The real (upper panel)
and imaginary part (lower panel) of a.c. susceptibility of \DTTO\
($x$ = 0, 0,5, 1.25, 1.5 and 1.8) at $H$ = 5 kOe. Pointed by the
arrows are: low temperature freezing peak (blue arrow), the
single-ion effect peak associated with \Dyion\ (green arrow) and
the \Tstar\ peak (red arrow). Insets in the $x$ = 0 plots are the
real (upper inset) and imaginary part (lower inset) of a.c.
susceptibility of \DTTO\ with $x$ = 0 at $H$ = 10 kOe.}

\end{figure*}

\begin{figure*}[t]
\centering \includegraphics[width=1.9\columnwidth]{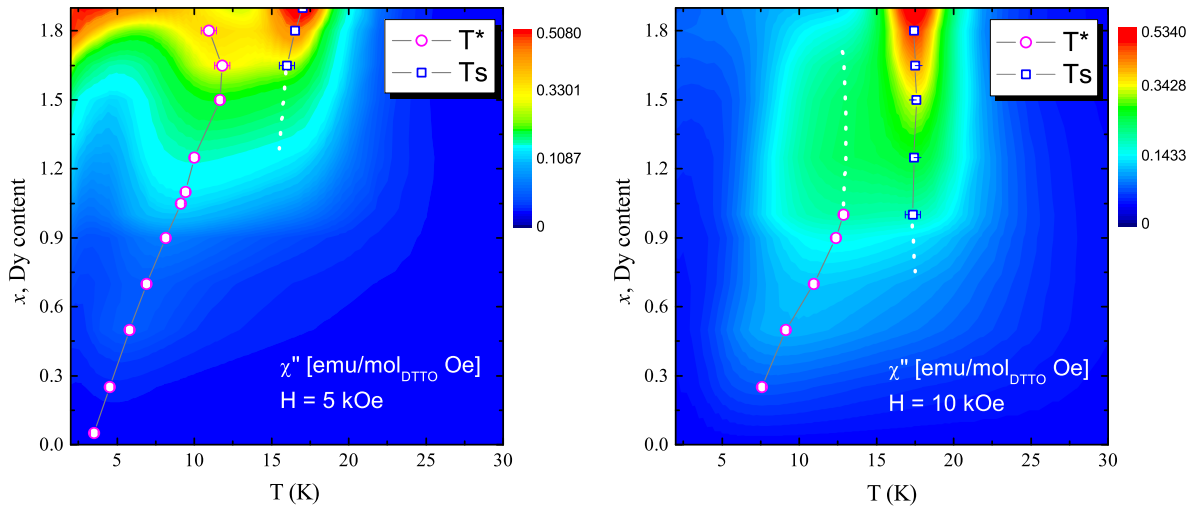}

\caption{\label{contour}(color online) Contour plot of \chiimag\ of
samples with different $x$ at $H$ = 5 and 10 kOe with $f$ = 10 kHz.
Pink circles (blue squares) denote \Tstar\ (\Ts ). Dotted lines are guide to
the eye where the peak positions are not well resolved. }
\end{figure*}

The two spin-freezing-like peaks cannot be attributed to a
collective spin glass transition, although the random occupancy of
rare earth sites by \Dyion\ and \Tbion\ is a likely source of spin
glassiness. In contrast to the spin glass transition, the d.c.
susceptibility of DTTO shows simply a Curie-Weiss behavior, more
strictly, a superposition of Curie-Weiss and Van Vleck
terms,\cite{Zinkin1996, Gingras2000prb, Ueland2006} with no freezing
transition. Secondly, the application of a magnetic field enhances
the characteristic temperature of the two freezing-like peaks, as
opposed to the case of spin glass where the transition temperature
is suppressed.\cite{Binder1986} Furthermore, the interactions for
Dy-Dy and Tb-Tb on the ground state doublet in DTO and TTO cases are
only of the order of 1 K, much lower than the temperature of the
observed peaks. Therefore the freezing-like peaks are most likely
not due to spin glass transitions. With higher magnetic fields ($H =
10$ kOe data shown in Fig. \ref{contour}(b) and higher field data
not shown here), the thermal spin contribution is suppressed at low
and high temperatures, leaving only a broad peak in \chireal\ as
observed in TTO and DTO.\cite{Ueland2006} Other samples with 0.5 $<
x <$ 1.5 show qualitatively similar behaviors. The appearance of the
two freezing-like peaks as well as their strong frequency dependence
are reminiscent of the single-ion effect peak in pure DTO, i.e., $x
= 2.0$, and will be discussed in detail by examining the composition
dependence below.

Fig. \ref{x-dependence} shows the a.c. susceptibility of five
representative samples with $x$ = 0, 0.5, 1.0, 1.5 and 1.8 at $H$
= 5 kOe. For $x$ = 0, there appears to be a peak below 3 K in
\chireal. The imaginary part of the sample with $x$ = 0 is clearly
different from other doped samples regarding their frequency
dependence. This is presumably related to the fact that the
exchange coupling and the energy gap between narrowly spaced CF
levels of TTO are of comparable energy scale and therefore both
have nonnegligible contributions to the
susceptibility.\cite{Gingras2000prb, Ueland2005} For samples with
$0 < x < 0.5$ (data not shown here), the crystal field
contribution increases progressively as $x$ increases, manifesting
as the transition to the behavior of the sample with $x = 0.5$.
The sample with $x = 0.5$ shows a single peak in \chiimag\
(\Tstar\ indicated by the red arrow). Additional peaks (\Tf\ and
\Ts\ indicated by the blue and green arrows) start to emerge for
$x$ = 1.0, and become more pronounced in $x$ = 1.5 and 1.8
samples. These additional peaks for $x \geqslant 1.0$ have the
same origin as the two peaks in pure DTO. In pure DTO, two peaks
are observed in \chiimag, a high temperature peak corresponding to
the single-ion effect, and a low temperature peak associated with
the spin freezing dictated by the ice-rule ground
state.\cite{Snyder2003, Snyder2004, Ke2007, Snyder2004b} For the
sample with $x$ = 1.8, the position of the high temperature peak
(indicated by the green arrow) are in good agreement with the
single-ion peak in DTO. Moreover, the frequency dependence of the
high temperature peak and the extracted energy scale corresponding
to the crystal field splitting are qualitatively similar to that
of pure DTO, as discussed below. We therefore identify the high
temperature peak as the single-ion peak of \Dyion\ spins. The low
temperature peak (indicated by the blue arrow) also possesses a
similar characteristic temperature with the spin freezing peak in
DTO,\cite{Snyder2004b} the corresponding peak in \chiimag\ should
appear below 2 K, the limit of PPMS, thereby only a rapid increase
in \chiimag\ upon cooling below 5 K is observable, and a direct
comparison with the DTO data is not allowed. Following the
convention of previous reports, we define the characteristic
temperature \Ts\ as the high temperature peak position (green
arrow) in \chiimag\, and similarly \Tf\ for the low temperature
peak (blue arrow).

The most important new feature in all samples studied, is the
presence of a third peak (indicated by the red arrow) between \Tf\
and \Ts. We define \Tstar\ ($T_{f}<T^{*}<T_{s}$) as the
characteristic temperature for this third peak. \Tstar\ peak
should root in the \Tbion\ spins as it exists at regions with
small $x$ (including $x = 0$ at $H = 10$ kOe, see the insets in
Fig. \ref{x-dependence}) , where the peaks associated with \Dyion\
disappear and the effect of \Dyion\ should play a minor role.
Combined with the susceptibility of
Dy$_{1.0}$Tb$_{1.0}$Ti$_{2}$O$_{7}$ at different fields shown in
Fig. \ref{field-dependence}, it is interesting to note that the
evolution of the a.c. susceptibility with increasing $x$ (Fig.
\ref{x-dependence}) is similar to that with constant $x$ but
increasing magnetic field (Fig. \ref{field-dependence}). In both
cases, the increase of $x$ or magnetic field leads to the
appearance of the \Tstar\ peak and subsequently the \Ts\ peak.

In order to investigate the origin of the \Tstar\ peak, we studied a
series of samples with $x \in$ {[}0, 2{]}, and track \Tstar\ and
\Ts\ in all samples when observable. A global view of \chiimag\ with
different $x$ is shown in the contour plot in Fig. \ref{contour}. It
is clearly seen that the \Tstar\ peak exists in a wide range of $x$.
Most interestingly, \Tstar($x$) increases linearly with increasing
$x$ in a wide range of $x$, both at 5 kOe and 10 kOe. At $H$ = 10
kOe, the low temperature upturn is strongly suppressed compared with
the 5 kOe case, as mentioned above. The linearity of \Tstar($x$)
persists in a wide $x$ range, from $x = 0$ to 1.5, obvious deviation
from the linearity occurs when $x > 1.5$. The characteristic peak
temperature of \Ts\ is also shown in Fig. \ref{contour}, with a
slight decrease as $x$ becomes smaller.

A clear deviation from linearity in \Tstar\ for $x \geqslant 1.5$ is
also an indication for its \Tbion\ origin. The pyrochlore structure
of DTTO consists corner-shared tetrahedrons, by assuming that
\Dyion\ and \Tbion\ spins in DTTO occupy the lattice randomly
therefore uniformly, there are two critical values of $x$(Dy) where
large change in local environment (i.e. tetrahedron that consists of
only one spin species: all \Dyion\ or all \Tbion) emerges: for $x >$
1.5 ($x \leqslant\ 0.5$), tetrahedrons of \Dyion -only (\Tbion-only)
spins appears, while \Tbion (\Dyion) still lies in the uniform
neighborhood coordinated with six \Dyion\ (\Tbion) spins. The
deviation from the linearity in \Tstar($x$) can be understood as a
direct consequence of the critical value of $x$ = 1.5. It is
important to note that there is no such deviation at the Tb end. The
presence of the deviation at $x = 1.5$ as well as the lack of that
at $x = 0.5$ is a strong indication of the asymmetric roles of
\Tbion\ and \Dyion\ spins in the origin of \Tstar\ peak, consistent
with our argument that the \Tstar\ peak is associated with \Tbion.

\begin{figure}[t]
 \centering \includegraphics[width=0.95\columnwidth]{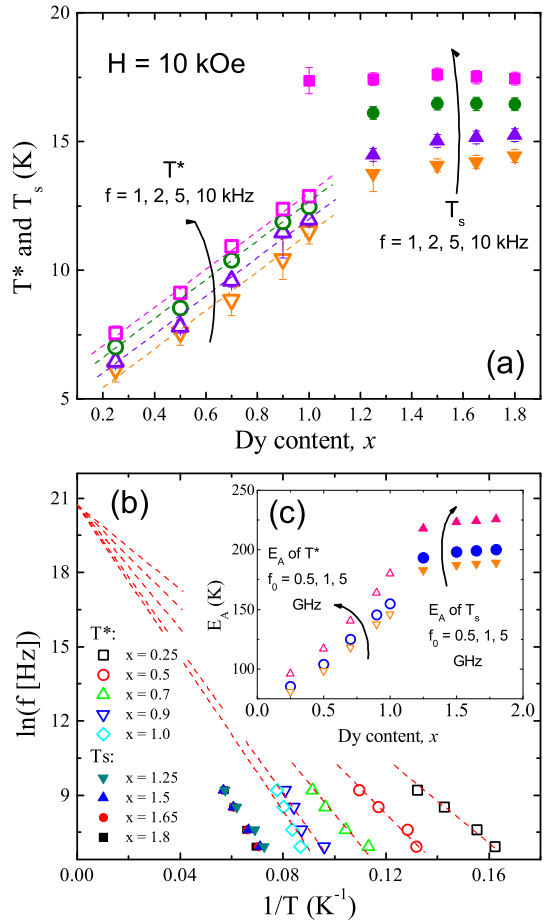}

\caption{\label{Ea-x}(color online) (a) At $H$ = 10 kOe, \Tstar\ (empty symbols) and \Ts\ (solid symbols) as a function of $x$(Dy) at different frequencies. Dash lines are guide to the eye to show the quasilinear increase of \Tstar. (b) Fittings by $f = f_{0}exp(-E_{A}/k_{B}T)$ with $f_{0}$ taken to be the physically reasonable value of 1 GHz. (c) The extracted energy barrier \Ea\ for \Tstar\ (open symbols) and \Ts\ (solid symbols) with fitting parameter $f_{0} = 0.5, 1$ and $5$ GHz.}
\end{figure}

To see the nature of \Tstar, we show in Fig. \ref{Ea-x}(a) the phase
diagrams of the \Ts($x$) and \Tstar($x$) with different a.c.
excitation frequencies (1, 2, 5 and 10 kHz) at $H = 10$ kOe. The quasilinear increase of \Tstar(x) is preserved at $H = 10$ kOe at different frequencies. The frequency dependence of both \Tstar\ and \Ts\ behaves according to the Arrhenius relation, as shown in Fig. \ref{Ea-x}(b). One note that to achieve a reliable fitting with the exponential function  $f = f_{0}exp(-E_{A}/k_{B}T)$, data with frequencies spanning over several decades are necessary. In our case, however, at frequencies lower than 1 kHz, \Tstar\ and \Ts\ peaks are indistinguishable, making the data at lower frequency range inaccessible. Thus we only focus on the data above 1 kHz. From the fitting of $f = f_{0}exp(-E_{A}/k_{B}T)$, where $f_{0}$ is taken to be the physically reasonable values of 1 GHz,\cite{Snyder2001,
Snyder2003} the energy barrier \Ea\ can be extracted. The extracted energy barrier \Ea\ is shown in Fig. \ref{Ea-x}(c), including \Ea\ by fitting with $f_{0} = 0.5$ and $5$ GHz to verify the reliability of the fittings. The validity of the fitting by Arrhenius law indicates that in the presence of a magnetic field the spin relaxation is thermally
driven.\cite{Snyder2003} In pure DTO, \Ea\ has been found to be
200 K (Ref.\cite{Snyder2004}). This energy barrier has been
interpreted as the gap between the lowest lying crystal field levels
of \Dyion. In DTTO, it can be seen that \Ea\ associated with \Ts\ changes only slighly from 193 K for $x$ = 1.25 to 200 K for $x$ = 1.8, a value very close to the energy barrier in pure DTO. This again proves our understanding that \Ts\ peak is associated with the single-ion effect of \Dyion. On the other hand, \Ea\ associated with \Tstar\ (attributed to \Tbion\ above) increases rapidly from 85 K for $x$ = 0.25 to 154 K for $x$ = 1.0. These values are unusually high, since the crystal field gap of ground state doublets for TTO is only O(20 K). Similar fittings with data at 5 kOe cannot be performed since the \Tstar\ and \Ts\ peaks are not well resolved for low frequencies.

Our observations including the field, composition and frequency
dependence of \Tstar\ raise the question of its underlying
mechanism. Knowing the exchange and dipolar interaction in both DTO
and TTO (for TTO, nearest neighbor exchange J$_{nn}$ = -0.88 K,
nearest dipolar interaction D$_{nn}$ = 0.8 K
(Ref.\cite{Mirebeau2006}), for DTO J$_{nn}$ = -1.2 K, D$_{nn}$ =
2.35 K (Ref.\cite{Hertog2000})), it is hard to imagine that certain
combinations of interactions between Dy and Tb spins (i.e. Dy-Dy, Tb-Tb and Dy-Tb)
can give rise to a characteristic temperature \Tstar\ with its
energy barrier two orders of magnitude higher than the energy scale
set by the interactions. Instead, the large energy gap extracted
from the frequency dependence of \Tstar\ suggests that single-ion
effects may play an important role. The CF levels for pure TTO and
DTO have been studied in detail previously.\cite{Rosenkranz2000,
Gingras2000prb, Gardner2001} For the DTTO samples, there exist two sets of CF schemes associated with \Dyion\ and
\Tbion\ respectively, with their CF level structures
determined by the local chemical environment. The zero-field \Ea\ in pure DTO is around 200 K. With $H$ = 10 kOe, \Ea\ associated with \Dyion\ in DTTO also stays around 200 K as is shown in Fig. \ref{Ea-x}(c). Thus one may expect that the single-ion peak temperature \Ts\ corresponding to the CF of \Dyion\ in DTTO should be robust against the change in magnetic field. However, according to the a.c. susceptibility data at zero field, e.g., Fig. \ref{field-dependence}(a$^{\prime}$ ) and (a$^{\prime\prime}$ ), none of the sample with $x \leqslant\ 1.5$ show signs of a single-ion peak above 2 K, implying narrowly spaced CF levels similar to that in TTO. Moreover, the field dependence of \Tstar\ and \Ts\ in Fig. \ref{field-dependence} indicate a dominant impact on their respective CF gap by the field. The possible field dependence of the CF levels is novel and deserves further investigation. Here with the current limited data, we discuss possible mechanisms that may give rise to the field dependence. It is known that with the application of a magnetic field, the Zeeman
splitting of the spin states may cause changes in the CF scheme.
Take \Dyion\ ions as an example,\cite{Fukazawa2002} $g_{J}J\sim$
O$(10)\mu_{B}/Dy$. With the magnetic field of 1 T, the Zeeman energy
is $g_{J}JH$ $\sim$ O(10) K, which is one order of magnitude smaller
than the experimental \Ea\ value. We therefore reason that there
should be some interactions beyond the CF interaction that
contribute to the magnetic field dependence. Previous studies on TTO
and similar geometrically frustrated systems revealed the softening
of the crystal field gap due to the presence of the spin-spin
correlation.\cite{Gardner1999, Zhou2008} The spin-spin correlation
can significantly alter the CF in TTO since its energy gap is only
O(10) K. It is therefore possible that the interplay between CF
interaction and spin-spin correlation dominate the magnetic field
dependence of the single-ion peaks. At zero field, the gap of CF is softened due to the spin-spin correlation between
\Dyion\ and \Tbion, causing the disappearance of the \Ts\ and \Tstar\  peak. With the presence of a field, the spin-spin correlation is dramatically altered, the gap of CF is again activated and thus the
peaks appear. However, the spin-spin interactions are orders of
magnitude smaller than the gap observed here, thus may not cause
sizable softening of the gap. Another possibility is the presence of
magneto-elastic coupling\cite{Alexandrov1985, Mirebeau2002, Mirebeau2004}. A giant magnetostriction has been reported in TTO (Ref.\cite{Alexandrov1985}), indicating strong coupling between spin and lattice degree of freedom. In \DTTO, it is possible that similar magneto-elastic coupling act in a way that the magnetic field may change the chemical environment of the
RE ion, and tune the crystal field effectively. However, such a strong effect on crystal field (change by several tens of Kelvin) by a moderate magnetic field (1 T) is still surprising. The detailed mechanism remains to be explored.

At $H$ = 10 kOe, with \Ea\ associated with \Dyion\ attaining its zero-field value of 200 K, the composition dependence of the \Tstar\ peak can be understood as a consequence of the change of the crystal field with \Dyion\ doping. The crystal field scheme is intimately related to the structural parameters of the system,\cite{Gingras2000prb} including the lattice constants and the positional parameters of the eight oxygen surrounding the RE ion. \Dyion\ doping leads to a systematic change in the structure and the corresponding changes in the crystal field levels. A very recent specific heat measurement on
the same system\cite{notes-0} shows that for small concentration of
Dy, a high temperature peak corresponding to the first excited CF
level of \Tbion\ moves to higher temperature, indicating a modification
of the CF of \Tbion\ site, which is consistent with our understanding.
Also, the linearity of \Tstar($x$) can be understood as a
consequence of the monotonic increase of $E_{A}(x)$ by the Arrehnius relation with suitable parameters.
Another possibility for the composition dependence of \Tstar\ peak
is the crystal field-phonon coupling as suggested in
TTO,\cite{Lummen2008} as \Dyion\ doping alters the phonon spectrum
and thus the CF level scheme.

The aforementioned similarity between the evolution of the a.c. susceptibility with increasing Dy doping $x$ and that with constant $x$ but increasing magnetic field now has a plausible interpretation. In both cases, the change in crystal field levels dominates the behavior of the a.c. susceptibility: in the former case, the Dy doping changes the structure and alters the crystal field directly; in the latter case, the magnetic field affect the crystal field possibly through the magneto-elastic coupling.

\section{conclusion}

To conclude, we have systematically studied the a.c. susceptibility
of \DTTO. A new peak associated with single-ion \Tbion\ (\Tstar\ peak) is observed, together with another peak associated with that of \Dyion\ (\Ts\ peak).
The phase diagram of both \Tstar($x$) and \Ts($x$) is presented showing the
characteristic temperature increasing linearly with $x$. Both \Ts\ and \Tstar\ peaks show strong frequency dependence, suggesting that the
transition is thermally activated. We interpret the origin of the
energy barrier to be associated with the gap in the lowest lying CF
levels of \Dyion\ and \Tbion. However, the extracted energy barrier for the \Tstar\ peak is an order
of magnitude higher than that of TTO and comparable to that of DTO. Furthermore, despite the large energy barrier, O(100 K), these characteristic temperatures are strongly influenced by applied magnetic field. Further neutron experiments and theoretical calculations may help to fully understand the system.

\begin{acknowledgments}
We gratefully acknowledge M. J. P. Gingras, P. Schiffer, Y. Zhou
and Y. Liu for very helpful and stimulating discussions. This work
is supported by the NSFC (Grant No.10634030), NSF of US (Grant No.
DMR0547036), National Basic Research Program of China (Grant No.
2007CB925001) and the PCSIRT of the Ministry of Education of China
(Grant No. IRT0754).
\end{acknowledgments}

\end{document}